\documentclass[letterpaper,10pt]{article} 
\usepackage{osameet3} %% use version 3 for proper copyright statement
\usepackage{amsmath,amssymb}
\usepackage[colorlinks=true,bookmarks=false,citecolor=blue,urlcolor=blue]{hyperref} %pdflatex
\usepackage{comment}

\usepackage[font=small]{caption}
\usepackage[font=footnotesize]{subcaption}
\begin{document}
\title{Real-Time Trans-Oceanic Fiber Sensing via Convergence of Long-Reach OFDR, High-Speed Optical Networking, and RDMA-Capable Edge-GPUs}

% No just change it!!! 
%% SUGGESTIONS
% Real-Time Transmission over Cabled Uncoupled and Coupled Core Multi-Core Fibers
% succes.................................................
% Real-Time MIMO TX in deployed strongly coupled multi-core fiber
% Field demonstration of real-time CC in italy
% Pizza n Pasta

\vspace{-0.5cm}
\author{
    M.~Mazur\textsuperscript{(1)}, 
    N.~K.~Fontaine\textsuperscript{(1)},      
    R.~Ryf\textsuperscript{(1)}, 
    S.~Bidkar\textsuperscript{(1)},
    T.~Chodkiewicz\textsuperscript{(2)},
    K.~M.~McBrian\textsuperscript{(3)}, \\
    K.~L.~McLaughlin\textsuperscript{(4)}, 
    B.~J.~Sperry\textsuperscript{(4)},
    J.~Rottman\textsuperscript{(3)},
    A.~G.~Butler\textsuperscript{(3)},
    M.~Karrenbach\textsuperscript{(4)}, \\ %artinkarrenbach@gmail.com     
    %V.~Kamalov\textsuperscript{(4)}, \\  %vkamalov@gmail.com 
    B.~Nguyen\textsuperscript{(1)}, 
    L.~Dallachiesa\textsuperscript{(1)}, 
    E.~Burrows\textsuperscript{(1)}, 
    D.~Winter\textsuperscript{(1)},     
    H.~Chen\textsuperscript{(1)},
    J.~Naik\textsuperscript{(5)}, 
    A.~Mistry\textsuperscript{(5)}, \\
    J.~Sandoz\textsuperscript{(2)}, 
    M.~Peterson\textsuperscript{(2)},
    M.~Libeskind\textsuperscript{(2)},
    B.~Le\textsuperscript{(2)},
    J.~Simsarian\textsuperscript{(1)} and    
    D.~T.~Neilson\textsuperscript{(1)}
}

\maketitle                  % Create title and author

%------------------------------------------ Description of Authors ----------------------------------------------%
%\vspace{4mm}
\address{
    \textsuperscript{(1)} Nokia Bell Labs, 600 Mountain Ave., Murray Hill, NJ 07974, USA.
    \textcolor{blue} {\underline{mikael.mazur@nokia-bell-labs.com}} \\
    \textsuperscript{(2)} Nokia, 600 Mountain Ave., Murray Hill, NJ 07974, USA. \\
    \textsuperscript{(3)} Leidos Inc., San Diego, CA 92121, USA \textsuperscript{(4)} Leidos Inc., Arlington, VA 22203, USA  \\
    \textsuperscript{(5)} Seismics Unusual, LLC, Brea, CA 92821, USA \\
    %\textsuperscript{(4)} Valey Kamalov LLC, Gainesville, FL 32607, USA\\
    \textsuperscript{(6)} Nokia Advanced Optics, 171 Madison Ave, New York City, NY 10016, USA \\
}
%\email{roland.ryf@nokia-bell-labs.com}
%%Uncomment the following line to override copyright year from the default current year.
\copyrightyear{2027}
 
%-------------------------------------------------- Abstract ---------------------------------------------------------%
\vspace{-6mm}
\begin{abstract}   
We demonstrate real-time network-integrated bi-directional fiber sensing over a 4400-km subsea cable. 
Streaming and fusing two OFDR instruments via RDMA over a live transoceanic 130\,Gbaud coherent link and GPU edge-compute enabling parallel capture of multiple spatial resolutions from $\approx$10\,m to $\approx$100\,m.
\end{abstract}

%-------------------------------------------------- Introduction Section -------------------------------------------------------%
\section{Introduction}
\vspace{-2mm}
\label{sec:intro}
The deployed telecom fiber grid could become a planetary-scale environmental sensor using distributed acoustic sensing (DAS), complementing sparse point sensors with continuous spatial coverage in areas such as the deep ocean, where seismic and tsunami monitoring remains scarce~\cite{lindsey2021fiber,lindsey2019illuminating,williams2019distributed,mazur2026high,10264088,mazur2025submarine,fernandezruiz2022seismic,ronnekleiv2024range,colares2026das,ip2022over}. 
DAS, especially transoceanic DAS, generates large quantities of data (TB/day), and often has to store it on a local hard drive, frequently in remote locations.
%%% NEED TO MAKE THE ONE SENTENCE TALKING MORE ABOUT THE DAS SYSTEMS
In parallel, we are witnessing the AI supercycle, where compute power and networking capacity is growing at incredible rate.  
For instance, a modest edge server with two NVIDIA RTX Pro 6000 GPUs and ConnectX-7 networking cards can support 800~Gb/s of bandwidth and 4~PFlops of compute, in addition to transporting multiple streams via remote-direct memory access (RDMA) over converged Ethernet (RoCEv2) between FPGAs and GPUs and the cloud.
One notable planetary compute and networking effort is the infiniCortex project~\cite{michalewicz2016infinicortex} which links supercomputers together over 100\,Gb/s links spanning continents.
With all this modern compute and networking combined with large datastreams produced by transoceanic DAS located across oceans, it makes sense to fuse numerous sensing interrogators (e.g., OFDRs) located around the globe, our global high-speed communication network, and massive compute nodes into a single global sensing system that can in real-time jointly process and improve the estimate of the environmental signals that modulate our cables.

Up to now, any multi-instrument or multi-location demonstration has relied on offline processing~\cite{biondi2023upper,mazur2026high}, where lack of integration with the telecom system, is a fundamental barrier for real-time network-scale sensing.

Here, we present a transoceanic bidirectional fiber-sensing field trial where the networking, coherent backhualing of real-time data, and joint processing of multiple OFDRs are all part of one single global sensing system.
% integrating transoceanic networking, multi-head fiber sensing and compute all into one single co-operating system. 
Two OFDR instruments one in Hawaii and the other in California, each stream $\approx$80\,Gbit/s of raw sensor data via RDMA over RoCEv2, mapped onto 100\,Gbit/s Ethernet and carried live within a 130\,Gbaud, 400\,Gbit/s coherent channel of a commercial (Nokia PSE-6) line system on the same 4400\,km cable. 
At the receiving end, a GPU edge server fuses both directions in real time, reducing the combined $>$160\,Gbit/s raw stream by $>$100$\times$ while pushing resolution to $\approx$10\,m --- $>$200,000 sensor points, again doubling our previous demonstration~\cite{mazur2026high}. 
Coherent processing of the full waveform lets a single acquisition generate multiple gauge lengths and sampling spacings in parallel, extending the dynamic range to capture weak seismic events alongside strong-motion signals without clipping, a key requirement for earthquake early warning. 
This inaugural demonstration of real-time multi-instrument fiber-sensing fusion carried over live coherent telecom transport, directly shows how sensing, transport, and edge/cloud compute can converge on shared subsea infrastructure.
\begin{figure*}[ht!]
   \centering
   \vspace{-3mm}
    \includegraphics[width=0.99\linewidth]{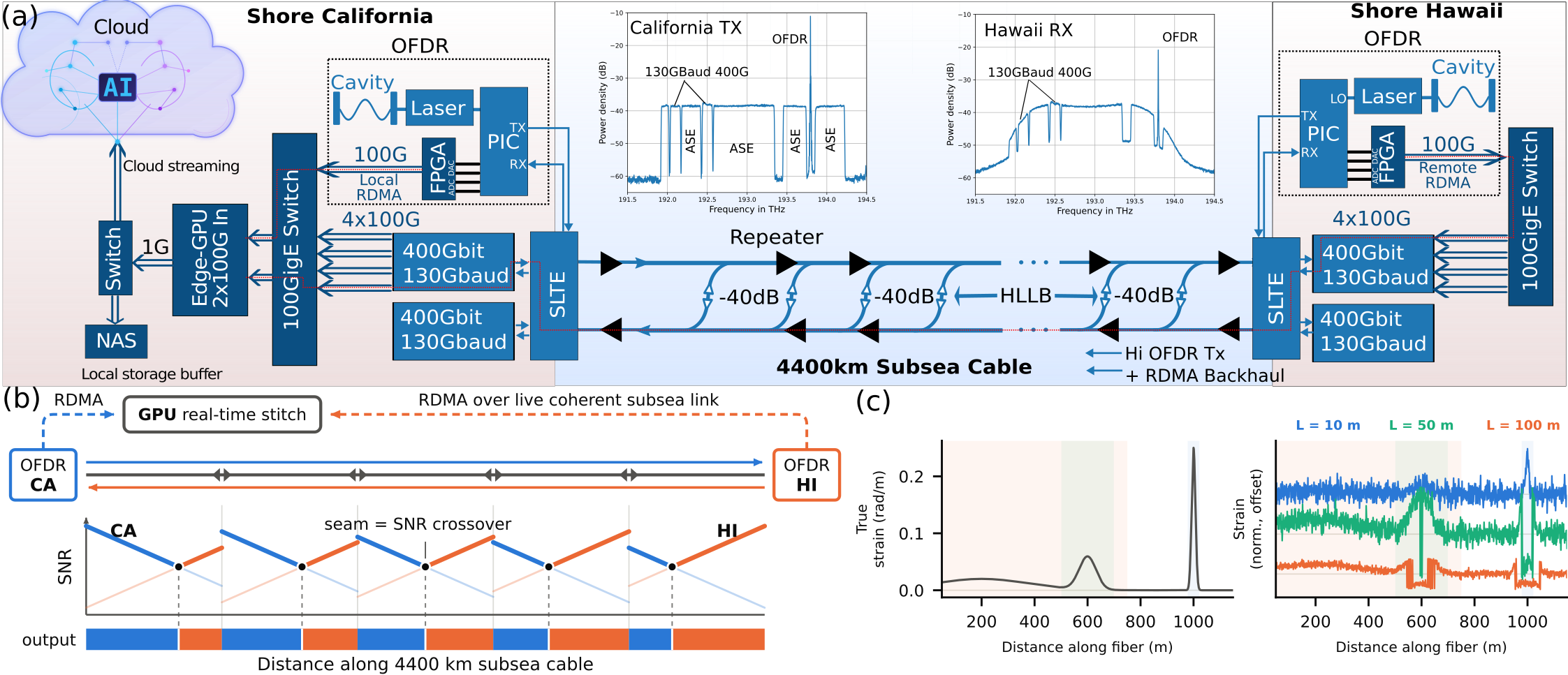}
    \caption{(a) Experimental setup showing the bi-directional sensing system implemented using real-time RDMA over a coherent subsea link connecting California (CA) and Hawaii (HI). 
    The two sensing streams are jointly processed by an edge-GPU server in California and the results streamed to the cloud. 
    The insets show the transmitted and received spectrum for the CA-HI direction. 
    Both data channels were running 400\,Gbit/s transmission. 
    (b) Schematic of the real-time SNR-aware bi-directional stitching process accounting for SNR degradation due to increased propagation distance. 
    (c) Illustration showing the limitation of processing using a single gauge length. To simultaneously support strong-motion monitoring for seismic early warning and high-sensitive weak signal detection, parallel processing is necessary. }
    \label{fig:setup}\vspace{-1mm}
\end{figure*}
%Mikael EDIT remaining, it's copy paste from OFC paper. 
% \vspace{-3mm}
\section{Experimental Setup}\label{sec:exp}
A sketch of the experimental setup is shown in Fig.~\ref{fig:setup}(a). 
It consists of four main components: subsea line terminal equipment (SLTE), a dual-channel 2$\times$1.2\, Tbit/s coherent optical line card (Nokia PSE-6 within Nokia's 1830 PSI-M DMAT6), and two long-reach OFDR sensors and a single streaming-capable GPU server. %the bi-directional fiber-sensing system implemented with two OFDRs. 
The $\approx$4400\,km, dispersion-managed subsea cable connects California (CA) and Hawaii (HI) via $\approx$100 inline optical repeaters ($\approx$45\,km span length) and offers $\approx$1.8\,THz of optical bandwidth. 
The SLTE uses an ASE source and a twin 1$\times$20 wavelength-selective switch (WSS) to combine channels and shape the ASE; passive couplers add the OFDR signals, in line with commercial SLTE implementations. 
The dual-channel PSE-6 transponder operated at 400\,Gbit/s per wavelength. 
Long-reach fiber sensing is enabled by the inline high-loss loopback couplers, adding $\approx$40\,dB of loss, conceptually illustrated in Fig.~\ref{fig:setup}(a). 
The CA-transmit and HI-received spectra are shown as insets in Fig.~\ref{fig:setup}(a). 
The main transceiver was placed at the edge of the band, suffering higher attenuation and $\approx$7\,dB of tilt across the channel width. 
The second channel is unused here but could instead be moved to validate co-existence between the sensing and data channels. 
The OFDR launch power was set to match the equivalent power of one open 100\,GHz slot; at this power, no BER penalty from the OFDRs was observed on the coherent transceivers and vice-versa -- the coherent transceivers did not degrade the OFDRs.

The CA and HI OFDRs each comprises a photonic integrated circuit (PIC), an FPGA with a 100\,Gbit/s Ethernet connection, and an NKT BASIK E15 laser locked to a vacuum reference cavity. 
Both used 250\,MHz chirp bandwidth and generated $\approx$80\,Gbit/s of raw sensor data, mapped directly into the shared GPU memory via ROCEv2. 
This relied on an IP network formed from two local 100\,Gbit/s networks (one per landing station), connected via the coherent trans-oceanic optical link. 
After joint processing on an edge GPU server in CA to combine the two sensor streams the output data volume was reduced by $>$100$\times$. 
The bi-directional stitching estimates the effective SNR of each direction along the cable and combines the two waveforms with per-span weights chosen to maximize the resulting SNR. 
This is conceptually illustrated in Fig.~\ref{fig:setup}(b). Since RDMA end-points can be placed anywhere, the GPUs could equally sit at the edge, in the cloud, or split across both, provided sufficient bandwidth is available. Importantly, the long round-trip delay of a trans-oceanic RoCEv2 link demands large buffers and an effectively lossless network --- avoiding packet loss is essential for trans-oceanic RDMA. The OFDR digital signal processing (DSP) followed~\cite{waagaard2021real}. To optimize performance over the entire cable and avoid clipping from phase wraps --- a known issue for DAS-based earthquake early warning --- we extended the DSP to process separate channel spacings (z-spacing) and gauge lengths (z-averaging) in parallel. This was enabled by combining a 100\%-duty-cycle OFDR with full-field coherent detection and DSP-based banding and filtering. Figure~\ref{fig:setup}(c) illustrates parallel processing across several gauge lengths from a single recorded trace.

\section{Results}\label{sec:results}\vspace{-1mm}
\begin{figure*}[ht!]
   \centering
   \vspace{1.5mm}
    \includegraphics[width=1.0\linewidth]{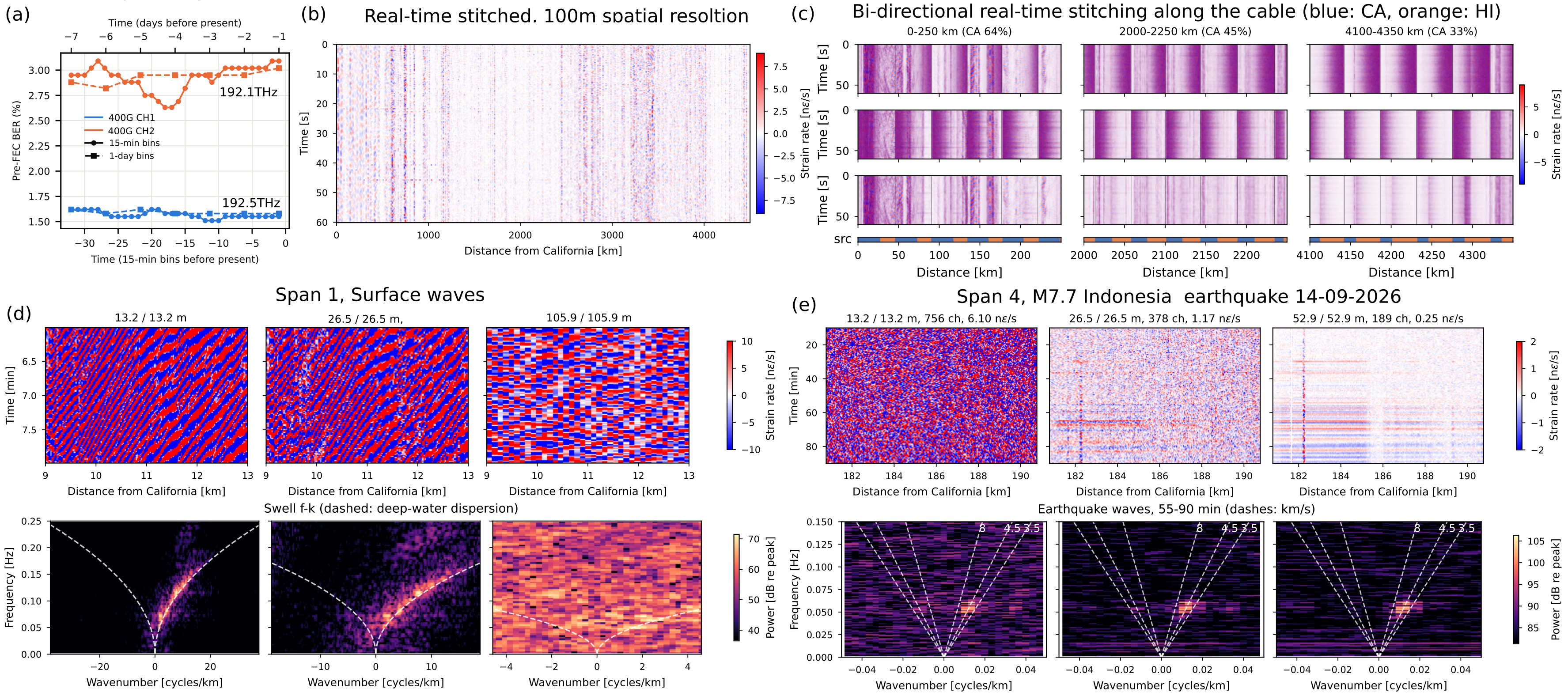}
    \caption{
    a) Transceiver Pre-FEC BER over 7 days (15-min and 1-day bins), showing stable performance. (b) 1-min real-time stitched waterfall at 100\,m resolution, showing ambient noise and cable-coupling variation along the full link. (c) SNR-aware stitching: combining both directions, accounting for propagation loss, maximizes the resulting SNR. (d) (top) Surface-wave measurements at three spatial resolutions/gauge lengths: clearly resolved at 13\,m but aliased at 26.5 and 100\,m. (bottom) Corresponding F-K plots show the surface-wave dispersion and the effect of spatial aliasing. (e) (top) Span-4 surface waves from the 2026-08-14 $M_w$7.7 earthquake after a 10--200\,mHz bandpass: the 13\,m data has insufficient SNR in the waterfall, but the signal is clearly visible at 100\,m. (bottom) F-K plots show the earthquake at all three spacings, confirming coherence.}
    \label{fig:result}\vspace{-1mm}
\end{figure*}
Using RDMA over trans-oceanic distances requires high-quality links. The bit error rate (BER) performance for both transceivers is shown in Fig.~\ref{fig:result}(a). Performance was measured in 15min bins and then combined into 1-day increments. The BER is stable for both transceivers, hovering around 3\% for channel 1 and 1.5\% for channel 2. No RDMA packet loss was observed. Note that channel 2 is carrying emulated data to allow performance monitoring. No penalty was observed on either transponder from adding the OFDR signals. Figure~\ref{fig:result}(b) shows a 1\,min waterfall with 100\,m spatial resolution after real-time stitching over the entire cable. The variation in strain rate is due to seafloor coupling and cable movements. A zoom-in on the stitching performance is shown in Fig.~\ref{fig:result}(c). The SNR-aware stitching resulted in relative contributions between 50\% and 66\%, depending on distance from shore. The ability to process multiple spatial resolutions is enabled by the real-time streaming using RDMA combined with GPU-based edge processing. The result is shown in Fig.~\ref{fig:result}(d) and (e) using surface swells near the California shore and a distant M7.7 Earthquake in Indonesia~\cite{usgs_us6000tkt2} as a source. Both waterfall plots and F-K diagrams are shown. For the surface swells in Fig.~\ref{fig:result}(d), we clearly see that to avoid spatial aliasing, 13.2\,m spatial resolution is needed. This is also highlighted by the clean F-K plot. However, the signal from the earthquake is weaker and the wavelength longer, pivoting for longer spatial resolution to increase the SNR. This is shown in Fig.~\ref{fig:result}(e). Importantly, all signals are processed at the same time using a single transmit and receive waveform, as discussed in the Experimental Setup section. The F-K plots in Fig.~\ref{fig:result}(e) also show the coherent nature of the fiber sensing array, with the earthquake source being visible on all spatial resolutions. 

\section{Conclusion}\label{sec:conclusion} \vspace{-1mm}
%We have demonstrated a converged sensing system that comprises transoceanic networking, two OFDR sensing heads located in California and Hawaii and a RDMA-Capable Edge-GPU located in California.
%The local OFDR RDMA stream transversed a local 100GigE network whereas the remote OFDR stream connected to this network via a 4400-km coherent line system.
%We intend to exploit the fast compute and networking advances seen in the AI supercycle to scale to a global system that jointly process sensing streams from cables around the world...  
By showing that sensing, transport, and compute can share one real-time coherent infrastructure, this work removes the offline-processing barrier that has confined fiber sensing to a handful of instruments at a time. Our demonstration over a 4400\,km California-to-Hawaii subsea cable combines the key building blocks -- long-reach OFDR-based fiber sensing, high-speed coherent transceivers, RDMA fabrics, and GPU compute -- each of which rides the same cost and performance curve as the broader AI build-out, giving a direct path to scale both coverage and performance. Extending this architecture across the global subsea network could turn today's telecom infrastructure into a single, continuously updating instrument with global coverage for earthquake, tsunami, and ocean-health monitoring.

% \section{References}

\vspace{-4mm}

%%%%%%%%%%%%%%%%%%%%%%%%%%%%%%%%%%%%%%%%%%%%%
%---------------------------------------------- End of Document -----------------------------------------------%

\vspace{-4mm}
\begin{thebibliography}{99} %% use BibTeX or add references manually
%\vspace{-1mm}
\footnotesize
\vspace{-2mm}
\bibitem{lindsey2021fiber} 
N. J. Lindsey and E. R. Martin, “Fiber-optic seismology”, Ann. Rev. of Earth and Planetary Sci., vol. 49, no. 1, pp. 309–336, 2021.

\bibitem{lindsey2019illuminating} 
N. J. Lindsey ~\textit{et al}, “Illuminating seafloor faults and ocean dynamics with dark...”, Science, vol. 366, no. 6469, pp. 1103–1107, 2019

\bibitem{williams2019distributed} 
E. F. Williams~\textit{et al}, “Distributed sensing of microseisms and teleseisms...”, Nat. Comm., vol. 10, no. 1, p. 5778, 2019.

\bibitem{mazur2026high} 
M. Mazur~\textit{et al}, “High-resolution trans-oceanic distributed acoustic sensing enabled by...”, Proc. OFC paper Th4C.7 2026.

\bibitem{10264088} 
Z. Wang~\textit{et al}, “Field trial of coexistence and simultaneous switching of real-time...”, JLT Vol. 42, No. 4, pp. 1304-1311, 2024.

\bibitem{mazur2025submarine} 
M. Mazur~\textit{et al}, “Submarine cable deep-ocean observation of mega-thrust earthquake and tsunami with”, Proc. ECOC 2025.

\bibitem{fernandezruiz2022seismic} 
 M. R. Fernandez-Ruiz~\textit{et al}, “Seismic monitoring with distributed acoustic sensing...”, JLT, vol. 40, no. 5, pp. 1453–1463, 2022.

\bibitem {ronnekleiv2024range} 
E. Rønnekleiv~\textit{et al}, Range-scalable distributed acoustic sensing with...”, Opt. Lett., vol. 50, no. 1, pp. 25-28, 2024.

\bibitem{colares2026das} 
R. A. Colares~\textit{et al}, “DAS over a 585-km EDFA-repeatered link with co-propagating...”, Proc. OFC paper M4J.6 2026.

\bibitem{ip2022over} 
E. Ip~\textit{et al}, “DAS over 1,007-km hybrid link with 10-Tb/s DP-16QAM co-propagation...”, Proc. OFC paper Th4A.2, 2022.

\bibitem{michalewicz2016infinicortex} 
M. T. Michalewicz~\textit{et al}, “Infinicortex: Present and future invited paper”, Proc. ACM, Int. Conf. Comp. Front. pp. 267-273, 2016.

\bibitem{biondi2023upper} 
E. Biondi~\textit{et al}, “An upper-crust lid over the long valley magma chamber”, Sci. Adv. Vol. 9, No. 42, 2023.

\bibitem{waagaard2021real} 
O. H. Waagaard~\textit{et al}, “Real-time low noise distributed acoustic sensing in 171 km...”, OSA Cont, Vol. 4, No. 2, pp. 688-701, 2021.

\bibitem{usgs_us6000tkt2} 
USGS, M 7.7 – 68 km NNW of Ende, Indonesia, USGS Earthquake Hazards Program. Event ID: us6000tkt2. Accessed: 17-08-2026.

%\bibliography{Mazur_References_Cleaned.bib}
% \bibitem{Kamalov2023}
% V. Kamalov,  “Methods for Geophysical Sensing on Submarine Cables, Proc. OFC paper W1H.1,  2023.

\end{thebibliography}
\end{document}